# AN HYBRID FRAMEWORK OTFS-OFDM BASED ON MOBILE SPEED ESTIMATION

Amina Darghouthi[1], Abdelhakim Khlifi[2], Hmaied Shaiek[3], Fatma Ben Salah[1] and Belgacem Chibani[1]

[1]MACS Laboratory: Modeling, Analysis and Control of Systems,
University of Gabes, Tunisia.
[2]Innov'COM laboratory, Sup'COM, University of Carthage, Tunisia.
[3]CEDRIC/LAETITIA Laboratory, CNAM, Paris, France.

## ABSTRACT

*The Future wireless communication systems face the challenging task of simultaneously providing high-quality service (QoS) and broadband data transmission, while also minimizing power consumption, latency, and system complexity. Although Orthogonal Frequency Division Multiplexing (OFDM) has been widely adopted in 4G and 5G systems, it struggles to cope with a significant delay and Doppler spread in high mobility scenarios. To address these challenges, a novel waveform named Orthogonal Time Frequency Space (OTFS). Designers aim to outperform OFDM by closely aligning signals with the channel behaviour. In this paper, we propose a switching strategy that empowers operators to select the most appropriate waveform based on an estimated speed of the mobile user. This strategy enables the base station to dynamically choose the waveform that best suits the mobile user's speed. Additionally, we suggest retaining an Integrated Sensing and Communication (ISAC) radar approach for accurate Doppler estimation. This provides precise information to facilitate the waveform selection procedure. By leveraging the switching strategy and harnessing the Doppler estimation capabilities of an ISAC radar.Our proposed approach aims to enhance the performance of wireless communication systems in high mobility cases. Considering the complexity of waveform processing, we introduce an optimized hybrid system that combines OTFS and OFDM, resulting in reduced complexity while still retaining performance benefits.This hybrid system presents a promising solution for improving the performance of wireless communication systems in higher mobility.The simulation results validate the effectiveness of our approach, demonstrating its potential advantages for future wireless communication systems. The effectiveness of the proposed approach is validated by simulation results as it will be illustrated.*

## 1. INTRODUCTION

Emerging wireless communication systems are designed to accommodate multiple waveforms, catering to a variety of mobility situations. Although, numerous wireless communication systems have made extensive use of Orthogonal Frequency Division Multiplexing (OFDM). However, it faces significant challenges in fastmovementenvironments. In such conditions, noticeable Doppler shifts and Doppler spread effects are usually observed. To address this issue, Orthogonal Time Frequency Space (OTFS), has been defined. This new waveform named OTFS takes advantage of delay and Doppler diversity. A superior performance over OFDM in high mobility contexts is registered. OTFS may be a promising candidate in this field due to its special waveform properties for high mobility wireless communication systems (HMWCS) [1], [2], [3]. For high mobility contexts, delay Doppler channel exhibits beneficial features like separability, stability, compactness, and possibly sparsity [4]. For future wireless systems generation named 6G, the significant challenge is the Doppler Velocity Estimation.The proposed 4G/5G technologies have introduced several enhancements for mobility scenarios. Indeed, with 4G mobile, handovers at speeds up to 350km/h can be performed with an allowable QoS [1], [7], [8], a higher mobility is a key performance for upcoming generation. Unfortunately, this technology presents sometimes interruptions causedto achieve higher transmission speeds for mobile terminals [2], [7]. To meet such goal, as in vehicle-to-everything (V2X), in drones, and in High Speed Rails (HSR).5G networks must support approaching 500km/h [2]. No later, for the frequency selective channels, one technique has considered or defining multi-carrier modulations (MCM) where action conducted on the frequency domain. With the upcoming availability of high mobility scenarios such as Hyper loop, future 6G is expected to support mobility at 1000km/h[2].High mobility induces significant Doppler shift and spread (i.e. the Doppler effect).Those imperfections appear directly in High Mobility Wireless Communications (HMWC) which suffer from rapid selective fading[3]. A compulsory role in communications is to look for matching the information to the propagation channel. Furthermore, the ingenious use of cyclic redundancy on transmission makes it possible to reduce terminalscomplexity. This is also empowered by Fast Fourier Transform FFT based algorithms usage.In 4G and 5G systems, processing methods were enhanced e.g. Orthogonal Frequency Division Modulation (OFDM) is becoming widely used as modulation structure for downlink communications. Data symbols has becoming multiplexed onto closely perfectly spaced orthogonal subcarriers. Even though, this waveform suffers from some limitations that making its main drawbacks. We can name e.g. high peak-to-average power ratio (PAPR), out-of-band (OOB) emissions, and significant loss of orthogonal waves in high mobility wireless channels [1], [2], [9]. Recently, a new bi dimensional (2D) waveform, named OTFS (Orthogonal Time Frequency Space), has been proposed [10], [11], [13] and [17]. One modulation's specificity is the usage of a pair of 2D transforms. This defines the known Symplectic Finite Fourier Transform (SFFT) and Inverse Symplectic Finite Fourier Transform (ISFFT) [4], [21]. In high mobility contexts, the OTFS systems achieve full diversity and greater performance compared to those obtained for OFDM [6], [7] and [27].Therefore, OTFS has received more attention. It is considered as a promising candidate for forthcoming generation of radio mobile networks [7], [18] and [19]. OTFS and OFDM waveforms both offer specific advantages and disadvantages tailored to varying mobility scenarios and system complexities. Interested to prove such merit and the improvements brought, we propose in this paper an original idea to define an alternate usage of such waveforms. It is noted that the OTFS is excellent for highmobility cases. However, it suffers from high processing complexity. In other side, OFDM is particularly well suited in low mobility situations. Consequently, this offers good performance and ease of use, but experiences a significant degradation in performance in faster moving situations [2], [7] and [14].Then, there is a dire need to find solutions that ensure high Quality of Service (QoS) simultaneously for different mobility rates. Currently, the use of an ISAC system for estimating various parameters, especially the speed of moving objects, is a promising approach for implementing OTFS and OFDM schemes. The goal is to achieve highly accurate estimates of delay, Doppler shift, object velocity, and target count. It is worth noting that most traditional velocity estimation methods rely on the delay Doppler (DD) technique. Several references, including [23], [24], [25],[28], mention radar integrated algorithms for this estimation. Really, users are practically, randomly distributed within the base station's coverage and they present varying mobility levels. The base station needs to select appropriate waveform to provide the best Quality of Service (QoS) offered for each user depending on their speed. Consequently, it becomes interesting to propose adequate solutions that provide adequate services simultaneously for both fast and slow speed moving users. When many users with varying mobility levels are randomly distributed within the base station's coverage area, the base station needs to select appropriate waveform to provide the best QoS offered for each user depending on this speed. In this paper, we propose a hybrid framework OTFS-OFDM based on mobile speed estimation. This estimation is carried out using a device that estimates the speed of objects more precisely, such as the ISAC radar. We have to recall that actions are empowered by radar ISAC system, which is based on the Matched Filter Fast Fourier (MF-F) algorithm. This algorithm is capable of obtaining an estimation of detection parameters with fractional precision, which improves the accuracy of the estimation. On the other hand, efficiency increases. Additionally, to reduce the number of comparisons needed in the search process, which speeds up the process and makes the algorithm more efficient. Let's note, the system performances depend strongly on such decision offering one usage among two possibilities named OFDM or OTFS. Ones the user's speed was estimated, we can see what will be speed value. This could be retained to switch between one of both strategies named OFDM or OTFS. This strategy could even more enhanced by defining a speed threshold value that we can define in order to operate the wanted selection. This arrangement is specifically designed to optimize the performance of OFDM over OTFS. To estimate the user mobility speed in order to assign that with the most matched waveform. After reviewing the aforementioned papers, the main contributions of this manuscript can be succinctly described as follows:

- Proposing a hybrid framework named OTFS-OFDM based on the speed estimation. This estimation is performed using a device that provides more accurate speed estimates of objects, such as the ISAC radar.

- Incorporating the ISAC radar sensing into the proposed framework is pivotal, particularly through the utilization of the Matched Filter-Fast Fourier (MF-F) algorithm. This algorithm, esteemed for its effectiveness, empowers the radar system to estimate detection parameters with fractional precision, thereby bolstering estimation accuracy.

- Measuring the probability of error of the proposed system, we find that the former OTFS under high mobility has a lower probability of error compared to OFDM. Furthermore, this strategy can be enhanced by defining a speed threshold value for selecting the desired strategy. This arrangement is specifically designed to optimize the performance of OFDM over OTFS. By estimating the user's mobility speed, we can assign the most suitable waveform, resulting in improved Quality of Service (QoS) and reduced complexity.

The remainder of the paper is structured as follows: In Section 2, we present a review of related work. Section 3 provides a brief overview of the structures of both OFDM and OTFS systems. Section 4 introduces the proposed framework and describes the speed estimation method utilized in our work. Moving on to Section 5, we present the simulation results of the proposed framework, along with a discussion on the system's complexity. Finally, in Section 6, we conclude the paper with a summary of our findings and outline potential avenues for future research.

## 2. RELATED WORK

Various research studies have been conducted on the vision and challenges of 6G technology [24], [31]. The objective of this research is to estimate various parameters used for waveform sensing. Several factors have been considered for the design of waveforms for integrated sensing and communication such as the (ISAC) system [28], [29], and [30]. Research topics worth exploring include wireless propagation path prediction and electromagnetic spectrum mapping [24], as well as, Terahertz technology [30]. The superior accuracy of ISAC estimation systems has led us to choose this system to estimate the velocity of moving objects. The authors of [26] introduce a two-dimensional radar imaging method using a MIMO OFDM radar, designed for automotive applications (the RadCom system was originally designed for use at 24 GHz). As its radar capability is comparable to that of conventional radar systems such as FMCW (frequency modulated continuous wave) radar, the authors aim to extend this capability to allow two-dimensional (2D) imaging, including range and azimuth, while maintaining speed estimation capability. Using receiver beamforming techniques and innovative radar processing methods, the paper [24] demonstrates the possibility of simultaneously estimating range, Doppler, and azimuth information for any number of objects, relative to the number of antenna elements in the array, during transmission.Among the various classes of antennabased on velocity (VE) estimation algorithms, MUSIC [24] is one of the most extensively studied. The MUSIC algorithm is easy to implement, with numerous versions that can be modified to fit different scenarios while providing high resolution. In [21], MUSIC was chosen because its spectrum can be directly represented as a radar image, without the need for post-processing of estimated object positions, unlike ESPRIT (estimation of signal parameters by rotational invariance techniques) [26]. The author in [22] addresses a critical challenge in the context of an integrated sensing and communication system (ISAC), namely improving the accuracy of the estimation of delay and Doppler shift parameters, essential parameters to support the performance of the communication system. To address this issue, the author presents a two-stage estimation algorithm known as the Fibonacci-matched filter (MF-F). This algorithm exploits waveform characteristics in orthogonal time-frequency space (OTFS) in the Doppler delay shift (DD) domain. For the first step (MF), the algorithm approximates the detection parameters by quantizing them on an integer grid, based on the relationship between the input and output signals of the ISAC model in the DD domain. This approximation is performed using the cyclic shift property of the matrix. In the second step (F), the author implements a twodimensional search technique based on the Fibonacci sequence, called the Fibonacci Search Method. This method provides an estimate of the detection parameters with fractional accuracy. It has the advantage of reducing the number of comparisons required and speeding up the search process. Finally, the author [25] propose a method using numerical simulations and hardware experiments. The results demonstrate that the MF-F method is capable of accurately estimating velocity and distance to the nearest millimeter, while exhibiting robustness and low complexity in numerical simulations. What's more, the Doppler shift and delay parameters estimated in the hardware experiments reach centimeter and meter levels. The author of [26] focuses on the field of integrated sensing and communication (ISAC), which is currently attracting a great deal of research interest. According to these estimation methods, the ISAC radar is the best for an accurate estimation of the parameters. This radar is used in our selection framework to choose between OTFS and OFDM.

## 3. SIGNAL MODELING

### 3.1. Basic OFDM Signal Modeling

Standard Data is transmitted using narrow subcarriers that make up the bandwidth. Each subcarrier transmits M-QAM symbols to an OFDM modulator. Although the transmission over the channel is successful, Inter Symbol Interference (ISI) often occurs. The modulation and demodulation processes can be performed by Fast Fourier Transform (FFT) and its inverse (IFFT) usage as illustrated in Figure 1. This issue is addressed by inserting a cyclic prefix (CP) between consecutive OFDM symbols. The channel delay spread is recommended to be longer than the CP's length to effectively mitigate ISI and simplify the equalization process [3],[5],[7],[8]. Subcarriers that are orthogonal help to enhance spectral efficiency.

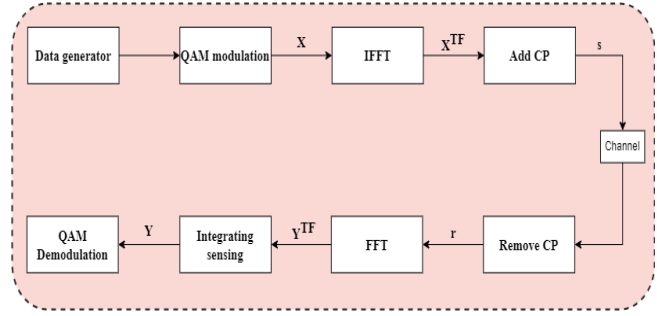

Figure1. OFDM Transmitter and Receiver Block Diagram

An OFDM system having respectively M subcarriers and N time slots is assumed. The total bandwidth of the OFDM signal is $B = M \Delta f$; with $\Delta f$ being the subcarrier spacing equal to 1. The frame duration will be $Tf = NT = NMT_s$, where T means a one OFDM symbol duration. This duration is equal to $MT_s$, where $T_s$ is the sampling time. We assume a static multipath channel with a maximum delay spread $\tau_{max}$ causing a channel Delay $\frac{\tau_{max}}{T_s}$. As stated before, to mitigate ISI and relax channel equalization task, the length of the cyclic prefix $L_{CP}$ should be greater than or equal to $l_{max}$, where $l_{max}$ that represents the maximum delay spread of the channel. In our case, we have chosen to take $L_{CP} = l_{max}$. The data symbols are defined as [5], [11], [13], [11]:

$$X[m,n] = m = 0, \ldots, M - 1; n = 0, \ldots N - 1 \quad (1)$$

Such data symbols are taken from the alphabet $A = \{a_1, \ldots, a_Q\}$, where $Q$ is the number of unique symbols in the alphabet ; and $\{a_1, \ldots, a_Q\}$: The individual symbols in the alphabet.

Each column of $X$ contains $N$ symbols. The transmitted signal can be expressed as follows [5], [7], [9], and [10]:

$$s(t) = \sum_{n=0}^{N-1} \sum_{m=0}^{M-1} X[m,n] g_{tx}(t - nT) e^{j2\pi m \Delta f(t-nT)} \quad (2)$$

where $g_{tx}(t) \geq 0$, for $0 \leq t < T$ is a pulse shaping waveform. We can define the set of orthogonal basis functions $\phi_{(n,m)}(t)$ used to shape M-QAM symbols as it follows [7,9,10]:

$$\phi_{(n,m)}(t) = g_{tx}(t - nT) e^{j2\pi m \Delta f(t-nT)} \quad_{,0 \leq m \leq M; 0 \leq n \leq N} \quad (3)$$

When the receiver is demultiplexing information, it utilizes the obtain basis signals. The process is described in [7, 9, 10] as follows:

$$\phi_{(n,m)}(t) = g_{rx}(t - nT) e^{j2\pi m \Delta f(t-nT)} \quad_{,0 \leq m \leq M; 0 \leq n \leq N} \quad (4)$$

where, $g_{rx}(t) \geq 0$ for $0 \leq t < T$ and is zero otherwise. This allows rewriting equation 2 as following [7], [9], [10]:

$$s(t) = \sum_{n=0}^{N-1} \sum_{m=0}^{M-1} X[m,n] \phi_{(n,m)}(t) \quad (5)$$

After that, a CP extension is then added to signal s (t) in order to overcome a multi- path channel effect. The cross-ambiguity function between the two signals $g_1(t)$ and $g_2(t)$ was defined as:

$$A_{g_1,g_2}(f,t) \triangleq \int g_1(t)g^*(t'-t)e^{-j2\pi f(t'-t)}dt' \qquad (6)$$

where defines the correlation between $g_1(t)$ and version of $g_2(t)$ delayed by t and shifted in frequency by f for all t and f in the time-frequency plane. When $s(t)$ passes through a time and frequency-selective radio channel, the received signal in the time domain is known as $r(t)$. let's $r(t)$ be the received time domain signal after propagation through a time- frequency selective wireless channel. This channel is characterized by its impulse response $h(t)$. so, the received signal can be expressed as [14], [15] and [16]:

$$r(t) = h(t) \circledast s(t) + w(t) \qquad (7)$$

where $w(t)$ is a complex Gaussian white noise.
To obtain, $r(t)$ is projected onto each $\phi_{(n,m)}(t)$. After the CP is removed, the received samples in Time Frequency (TF) applies FFT operator can be expressed as [14],[15],[16]

$$Y(f,t) = A_{g_1,g_2}(f,t) \triangleq \int g_{rx}^*(t'-t)e^{-j2\pi m\Delta f(t'-t)}, \qquad (8)$$

$$Y(m,n) = Y(f,t)_{f=m\Delta f, t=nT}. \qquad (9)$$

## 3.2. OTFS System Modeling

In this section, we will discuss the renowned concept proposed by Hadani, the OTFS approach [4], [11], [13]. More specifically, the system model associated with the OTFS scheme is illustrated in Figure 2.

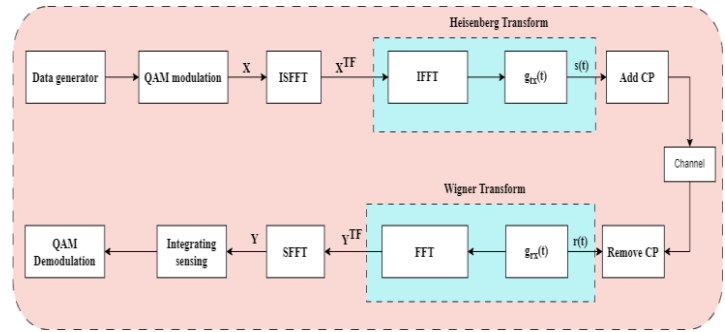

Figure 2. OTFS Transmitter and Receiver Block Diagram

OTFS is applied by mapping the previously prepared QAM symbols onto the delay-Doppler domain (DD). For that, the symbols are initially converted from the delay-Doppler domain (DD) to the time-frequency domain (TF). Firstly, at the transmitter, the QAM symbols are arranged in a two-dimensional (2D) matrix with N columns in the Doppler domain and M rows in the delay domain. The time-frequency grid is discretized to a $M$ by $N$ grid (for some integers $N, M > 0$), using intervals of T (seconds) and $\Delta f\ (Hz)$, the time and frequency axes are sampled, respectively, i.e., [4], [11], [12], [13]:

$$\Lambda = \{(nT, m\Delta f)\}, n = 0,\cdots,N-1, m = 0,\cdots,M-1\} \tag{10}$$

The modulated time frequency samples $X[n,m], n = 0,\ldots,N-1, m = 0,\ldots,M-1$ are transmitted over an OTFS frame with duration $T_f = NT$ and occupy a bandwidth $B = M\Delta f$. The delay Doppler plane in the region $(0,T] \times (\frac{-\Delta f}{2}, \frac{\Delta f}{2})$ is discretized to an $M$ by $N$ grid [4], [17], [18] and [19]:

$$\Gamma = \{(\frac{k}{NT}, \frac{l}{M\Delta f})\}, k = 0,\cdots,N-1, l = 0,\cdots,M-1\} \tag{11}$$

where $(\frac{k}{NT}, \frac{l}{M\Delta f})$ represent the quantization steps of the delay and Doppler frequency axes, respectively. Then the signal is transformed into the time-frequency domain through the inverse symplectic finite Fourier transform (ISFFT) in the second step. This will be written like [17], [18] and [19]:

$$X^{TF}[m,n] = \frac{1}{\sqrt{MN}} \sum_{n=0}^{N-1} \sum_{m=0}^{M-1} X[l,k] e^{j2\pi(\frac{nk}{N} - \frac{ml}{M})} \tag{12}$$

where $X[l,k]$ is the delay Doppler signal modulated pulse.

Each data frame in this scenario has a total frame duration of $B = NT$ and a bandwidth of $Ts = N\Delta f$. After reshaping the matrix $X^{TF}[m,n]$ into a time frequency domain sequence, the transmitted OTFS signal, denoted $s(t)$, can be derived by applying the Heisenberg transform to $X^{TF}$ with the transmitter shaping pulse, $g_{tx}(t)$. More specifically, the Heisenberg transform can be viewed as a multicarrier modulator. This Heisenberg approach involves using the conventional OFDM modulator. In particular, with conventional OFDM modulation, the Heisenberg transform could be achieved by an inverse fast Fourier transform (IFFT) module and transmit pulse shaping. In this scenario, the transmitted signal $s(t)$ using Heisenberg Transform as proposed by [11], [17], [18] and [19].

$$s(t) = \sum_{n=0}^{N-1} \sum_{m=0}^{M-1} X^{TF}[m,n] g_{tx}(t - nT) e^{j2\pi m\Delta f(t - nT)} \tag{13}$$

where, $g_{tx}(t)$ is the window function. It has been shown in [17].

Practical rectangular transmit and receive pulses are used, which are compatible with OFDM modulation. Finally, a CP is added to the time domain signal for every data frame, as indicated by [14].

$$S_{CP}(t) = \begin{cases} s(t) & 0 \le t \le T_s \\ s(t + T_s) & -T_{CP} \le t < 0 \end{cases} \tag{14}$$

where $T_{CP}$ denotes the duration of the CP.

The channel impulse response in DD is characterized by the target detection channel or communication paths transmitted. We suppose that the P multipath components, where i[th] path linked to complex path gain $\alpha_i$, delay $\tau_i$, and Doppler shift $v_i$. Where $\tau_i \in [0, \frac{1}{\Delta f})$, $v_i \in [-\frac{1}{2T}, \frac{1}{2T})$. In this situation, any two paths are solved in the delay Doppler domain (i.e., $|\tau_i - \tau_j| \ge \frac{1}{M\Delta f}$ or $|v_i -$ $v_j| \geq \frac{1}{NT}$. Therefore, the impulse response of the wireless channel in the DD domain is given as [30]:

$$h(\tau, v) = \sum_{i=0}^{P} \alpha_i \, \delta(\tau - \tau_i)\delta(N - N_i) \qquad (15)$$

For joint radar ISAC integrating sensing and communication, the delay and Doppler shifts are calculated using $\tau_i = \frac{r_i}{c_0}, v_i = \frac{f_c v_i}{c_0}$ where distance $r_i$ and velocity $v_i$ along the i$^{th}$ path, and $f_c$ is the carrier frequency and the speed of light is therefore represented by $c_0$. The integration of system detection requires consideration of both the round-trip delay and the Doppler effect. The calculations above have an extra multiplier of 2 added. For this instance, the path delay and Doppler shift correspond to integer multipliers of delay and Doppler resolution, $\tau_i = \frac{l_i}{M\Delta f}$ and $v_i = \frac{k_i}{NT}$. During the transmission, a signal can thus suffer from various changes, particularly in scenarios involving high mobility. These changes produce shifts both in the time and frequency domains. In these conditions, the received signal could be expressed as [17], [18] and [19].

$$r(t) = \int \int h(\tau, v) s(t - \tau_i) \, e^{j2\pi v(t - \tau_i)} d\tau dv + w(t) \qquad (16)$$

The received symbols matrix $Y^{TF}[m,n]$ in the Time Frequency Domain, is obtained by sampling the cross - ambiguity function $A_{g_{rx},r}(t,f)$ according to [17], [18],[19]:

$$Y^{TF}[m,n] = A_{g_{rx},r}(t,f)_{t=nT, f=m\delta f} \qquad (17)$$

where the sampling cross ambiguity function $A_{g_{rx},r}(t,f)$ as indicated:

$$A_{g_{rx},r}(t,f) \triangleq \int g_{rx}^*(t - t') r(t) \, e^{j2\pi(t-t')} \qquad (18)$$

Finally, the DD domain samples are obtained by applying the SFFT to $Y[l,k]$[4], [17]:

$$Y[l,k] = \frac{1}{\sqrt{MN}} \sum_{n=0}^{N-1} \sum_{m=0}^{M-1} X[m,n] e^{j2\pi(\frac{nk}{N} - \frac{ml}{M})} \qquad (19)$$

## 4. PROPOSED MODELLING

In this section, we will show how someone could adequately process the studied signal based on a chosen processing strategy. This will concern the ISAC radar's approach for estimating various parameters. This will obviously help to estimate the unknown speed characterizing a mobile user. Let's see how this will be done. As shown in Figure3, the Framework that associated with a signal processing system. This system comprises three main processing blocks. In the first place, we have the base station that including elements like Random Data, Inverse Fourier Transform Symplectic (ISFFT), and Heisenberg Transform. A transmitted signal is sent by this base station to the Sensing Target. Such target is a moving object like vehicles. An echo signal is returned to the receiver of the base station. This defines an integrated ISAC device. The ISAC receiver processes that signal using various tools like the Wigner transform, Fourier Transform Symplectic (SFFT), and the Sensing Signal Detection. The Sensing Signal Detection estimates various parameters, like the speed of the objects already detected by the Sensing Target. The signals are processed using based on the estimated speed of the objects and then comparing them to predefined a speed threshold [2], $v_{hreshold}$= [120kmh, 250kmh, 500km/h]. The adopted threshold will be so useful in order to split two speed ranges named low and high speed. When the estimated speed is below the threshold, the Sensing Signal Detection returns a signal to the base station ordering to complete the OFDM signal processing tasks. Otherwise, the speed will be above the threshold and that results allows retaining OTFS signal processing. As briefly explained, we see that the OFDM method concerns a low mobility situation.

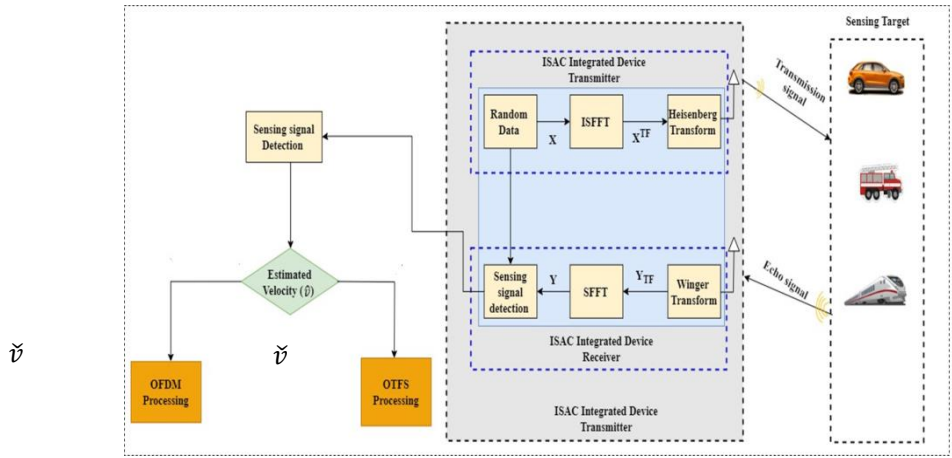

Figure 3. Framework of Hybrid OTFS-OFDM system

However, the OTFS approach is applied in High mobility conditions. The former analysis was so convenient to guide our sight to suggest a framework defining our contribution. The focus is on helping to develop a hybrid OTFS-OFDM system that is based on user estimation. This estimate is based on the estimate speed approach employed by the ISAC radar application. Before that let's give the principle of ISAC technique. In fact, the ISAC radar is based on the Matched Filter Fast Fourier MF-F algorithm. This algorithm has a significant impact on the improvement of detection parameter estimation in radar ISAC (Integrated Sensing and Communication). Indeed, this system has several advantages thanks to, a precision is enhanced by the use of the Fibonacci sequence. The MF-F algorithm is able to obtaining an estimation of detection parameters, with fractional precision. This improves the accuracy the estimation. On the other hand, efficiency increases. In fact, the MF-F algorithm reduces the number of comparisons needed, making the algorithm more efficient. In addition, the MF-F algorithm has demonstrated robustness in numerical simulations, which means it can provide accurate estimates even under difficult conditions. Finally, compared to other estimation algorithms that may have high complexity, the MF-F algorithm has relatively low complexity, which makes it easier to implement and use. Let's note that, the system performances depend strongly on such decision offering one usage among two possibilities named OFDM or OTFS. Ones the user's speed was estimated, we can see what will be speed value. The use of the MF-F algorithm to estimate velocity and detect data is crucial. The Doppler estimation, assisted by ISAC radar, is enhanced by an iterative refinement process, which guarantees higher accuracy and improved data detection. For that, the receiver processing allows us to implement two modes for ISAC radar applications: active detection, joint passive detection. These two modes have objectives that are described as follows [34]:

- The objective of active detection is to calculate channel delay $\tau$ and Doppler shift by considering transmit vector X and receive vector;

- The objective of passive detection and joint data detection is to estimate the channel parameters ($\alpha$, $\tau$, $\nu$) and recover X and the received vector Y. All previous indication described in [33].

57

International Journal of Computer Networks & Communications (IJCNC) Vol.16, No.3, May 2024

$$(\hat{\tau}_i, \hat{v}_i) = \arg max_{(\tau,v)\epsilon\Lambda_i}|(\Gamma_i)^H Y_i|^2 \quad (20)$$

where $\Gamma_i$ is the estimated Doppler and $c_0$ the velocity of sight and $f_c$ the carrier frequency. We can present firstly the estimated delay and Doppler based on the indication expression [33]. Where $\Gamma_i$ delay Doppler plane, the estimated velocity is indicated as follows [34]:

$$\hat{v}_i = \frac{\hat{v}_i c_0}{2f_c} \quad (21)$$

where $\hat{v}_i$ is the estimated Doppler and $c_0$ the velocity of sight.

## 5. RESULTS AND DISCUSSIONS

In order to check the previously described idea, we have chosen to make simulation under the conditions as summarized in the Table 1.

Table 1. Simulation Parameters of OTFS and OFDM.

| Parameter | Value |
|---|---|
| Channel Power Delay Profile | EVA |
| Subcarrier Spacing Δf | 15 KHz |
| Number of symbols per frame | 8 |
| Number of Subcarriers per Block | 16 |
| Carrier Frequency $f_c$ (GHz) | 0.95 |
| Velocity Estimation $\hat{V}$ (km/h) | 3,10,30,200,500 |
| Modulation | 4 − 16QAM |

### 5.1. Evaluation Performance of ISAC System

Our radar ISAC speed estimation system has been subjected to comprehensive testing to assess its capabilities. The outcomes underscored several significant attributes:

- **Enhanced Precision**: The ISAC system exhibited remarkable precision in estimating speed across diverse scenarios. For instance, in a controlled setting where a vehicle maintained a steady speed of 200 km/h, the system's speed estimation deviated by less than 2% on average. This degree of precision significantly surpasses that of traditional systems, which typically have an error margin about 10%.
- **Swift Response Time**: The response time of ISAC, defined as the duration from receiving input data to delivering a speed estimate, was impressively swift. On average, the system furnished a speed estimate in under 0.5 seconds. This rapid response time ensures the system's effective deployment in realtime applications.
- **Robustness**: We observed that the radar ISAC system demonstrated exceptional robustness, even under challenging conditions. For example, in situations of poor visibility or inclement weather, the system's performance remained stable.
- **The accuracy of speed estimates** under these conditions was on par with those achieved under optimal conditions. When juxtaposed with other speed estimation systems, our ISAC system excelled in terms of accuracy. Additionally, our radar ISAC system showcased superior robustness, sustaining high performance even under unfavorable conditions.

International Journal of Computer Networks & Communications (IJCNC) Vol.16, No.3, May 2024

To conclude, the results suggest that our ISAC system for speed estimation exhibits outstanding performance across various metrics. Its high accuracy and robustness under challenging conditions render it a valuable asset for a multitude of applications as illustrated in Figure 4.

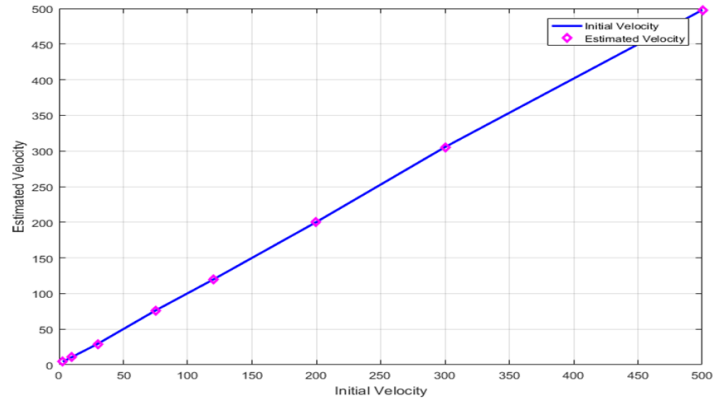

Figure 4. Performance Evaluation for ISAC Speed Estimation System.

## 5.2. BER Performance

This section evaluates the BER performance of the proposed OTFS-OFDM method using different speeds. Firstly, we consider an OFDM system as a function of estimated speed. Figure 5 and Figure 6 show the BER of the OFDM system with various speeds. The modulation schemes are respectively 4-QAM for Figure 5 and 16-QAM for Figure 6.

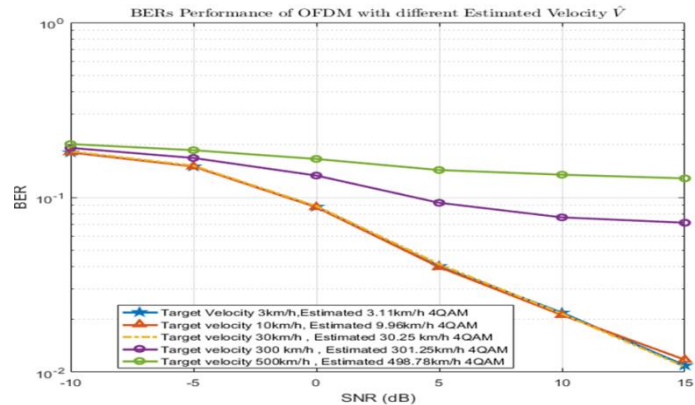

Figure 5. BERs Performance of OFDM: The modulation scheme is 4-QAM.

As shown in Figure 5, we have chosen five values for the speed estimate, e.g. (3km/h, 10km/h, 30km/h, 200km/h, 500 km/h). When the value of the speed estimate (3km/h, 10km/h, and 30km/h) is low, these cases have similar BER values. When the speed estimate is increased to 200 km/h and 500km/h, we find that the BER is the highest among those values. Furthermore, we find that the BER for high speed increases, as the signal power allocated to the channel speed causes inaccurate channel estimation and failure data detection.

59

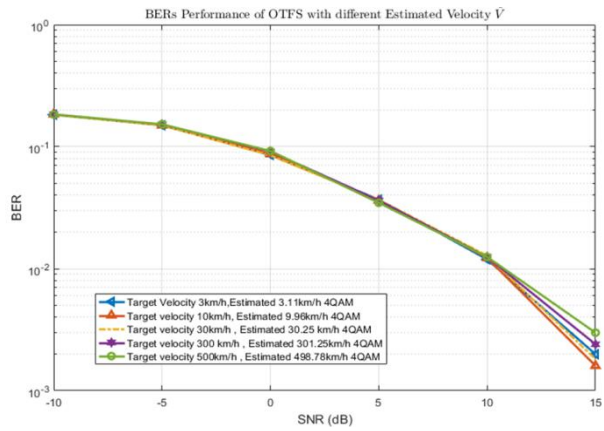

Figure 6. BERs Performance of OFDM: The modulation scheme is 16-QAM.

In Figure 6, we evaluate the BER from 0 dB to 15 dB for the 16-QAM modulation scheme, which requires a higher SNR to obtain a good BER. We observe that OFDM at low speed obtains the best BER, while OFDM at high speed obtains the lowest BER value at 0 dB to 15dB. For that, we can use the low speed for OFDM processing, which corresponds well to the simulation results in Figure 6. The results prove how in case of a low speed, OFDM approach give nearly optimal BER which's insensitive to used speed. Based on these results dealing with OFDM usage, we can conclude that the BER remains acceptable for lower speeds. However, when the speed increases, the BER shows a great increase causing bad performances. This inadequate choice in terms of processing tool must be revised to suggest a better tool to overcome such OFDM cons.We can conclude hat at low speeds, orthogonal frequency division multiplexing (OFDM) approach delivers a near optimal bit error rate (BER), which is relatively insensitive to the speed used. We can deduce from these results that BER remains within acceptable limits for low speeds when using OFDM. Whereas, at higher speeds, BER increases significantly, leading to limited performance. This highlights the need for a more appropriate processing tool to mitigate the limitations of OFDM at higher speeds. Furthermore, by solving the problem of the high mobility of this waveform, we can propose another optimal solution for modulating the average waveform. We find that the optimal solution for the proposed waveform is the OTFS waveform. Whether the SNR, the estimation of speed assisted by ISAC radar is less affected. Therefore, the higher the noise level, the more important it is to use an ISAC radar for accurate speed estimation.

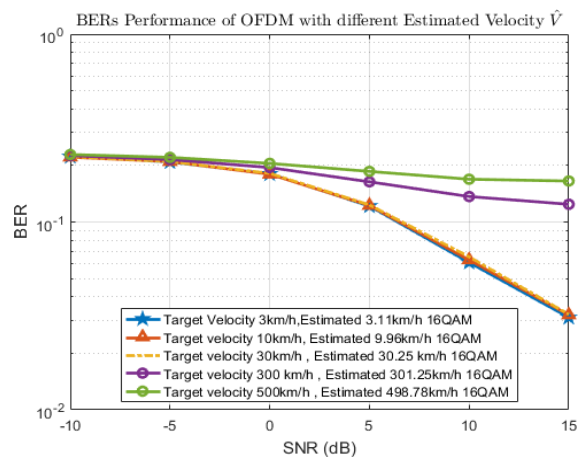

Figure 7. BERs Performance of OTFS: The modulation scheme is 4-QAM.

Figure 7 and Figure 8 clearly show that for a given SNR value, BER increases with estimated speed. This is a common observation when analyzing the performance of communication systems at different speeds. The modulation schemes are respectively 4-QAM for Figure 7 and 16-QAM for Figure 8.As the estimated speed increases, the BER increases accordingly, indicating a deterioration in system performance. As shown in Figure 8, since the higher order modulation scheme requires a higher SNR to obtain a good BER, we evaluate the BER from 0 dB to 15 dB, when the modulation scheme is 16-QAM. We observe that both low and high rate OTFS achieve the best BER.

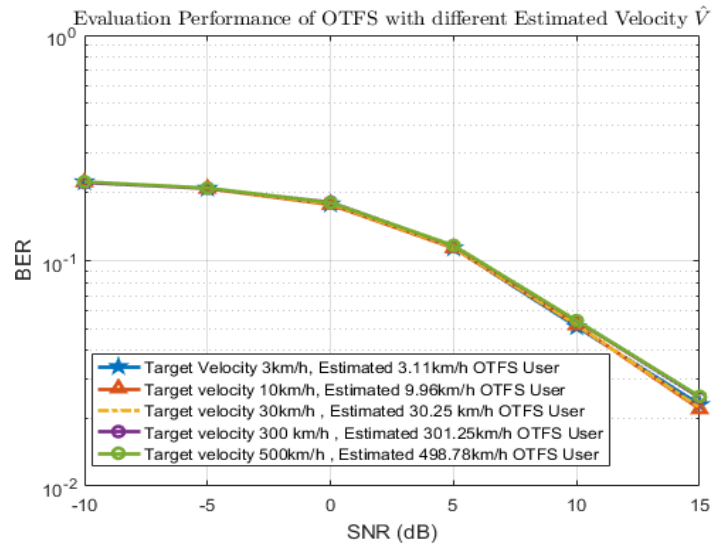

Figure 8. BERs Performance of OTFS: The modulation scheme is 16-QAM.

The following Figure 9 and Figure 10 illustrate the performance in terms of BER of a hybrid scheme for OTFS-OFDM systems. The modulation schemes are respectively 4-QAM for Figure 9 and 16-QAM for Figure 9. In Figure 10, this which combines the strengths of both waveforms to improve BER in high mobility scenarios. In parallel, we can use the low rate for OFDM processing. In addition, the high mobility problem is solved. We found that the OTFS filter is better because it is less noisy than the speed of the moving object. All these notes can lead to choose OTFS in high speed scenarios. Even that this technique could be applied for low speeds, its processing complexity compared to that OFDM, goes for OFDM retention due to its simple implementation. This phenomenon can be attributed to the Doppler effect, which induces changes in signal frequency and phase at higher speeds. These curves clearly show the adequacy of OTFS in case of high speed. No performance degradation can be obtained in such conditions when OTFS is used.

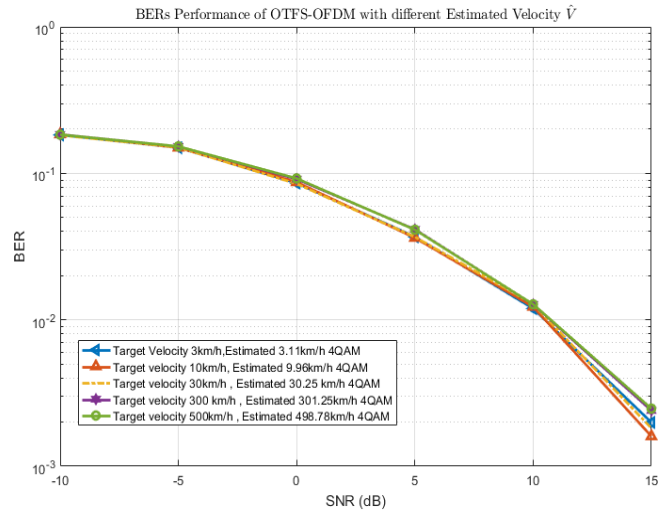

Figure 9. BERs Performance of OTFS-OFDM: The modulation scheme is 4-QAM

All these notes can lead to choose OTFS in high speed scenarios. In parallel, we can use the low rate for OFDM processing, which corresponds well with the simulation results in Figures. 9 and 10. In addition, the high mobility problem is solved. We found that the OTFS processing is better because it is less noisy than the speed of the moving object. All these notes can lead to choose OTFS in high speed scenarios. Even that this technique could be applied for low speeds, its processing complexity compared to that OFDM, goes for OFDM retention due to its simple implementation. This phenomenon can be attributed to the Doppler Effect, which induces changes in signal frequency and phase at higher speeds. These curves clearly show the adequacy of OTFS in case of high speed. No performance degradation can be obtained in such conditions when OTFS is used. All these notes can lead to choose OTFS in high speed scenarios. Even that, this technique could be applied for low speeds its processing complexity compared to that OFDM. However, despite its potential application at low speeds, the processing complexity of OTFS compared with OFDM often leads to OFDM being chosen because of its simpler implementation. The interesting fact is that the BER curves for all scenarios merge, indicating insensitivity to user speed. Such a feature underlines the relevance of OTFS in high-speed scenarios. Noteworthy,the data show that a high mobility user obtains a BER nearly similar to that of a low mobility. Even that, this technique could be applied for low speeds its processing complexity compared to that OFDM. However, despite its potential application at low speeds, the processing complexity of OTFS compared with OFDM often leads to OFDM being chosen because of its simpler implementation. The interesting fact is that the BER curves for all scenarios merge, indicating insensitivity to user speed. Such a feature underlines the relevance of OTFS in highspeed scenarios. The BERs of both OTFS and OFDM waveforms show significant variations, suggesting that the choice of waveform could be guided by radar ISAC based on the speed estimation. In particular, this approach can improve BER for high mobility users when using the OTFS waveform. On the other hand, for low mobility users, the OFDM waveform seems to be a reliable choice. The results show that the hybrid scheme offers better performance than using OTFS or OFDM waveforms alone. In low mobility scenarios, the BER of a hybrid scheme is similar to that of the OFDM waveform usage because the Doppler spread can be neglected, making the use of OTFS waveform being less advantageous. However, it is crucial to note that the performance of the hybrid scheme is heavily reliant on selecting appropriate parameters, such as the subcarrier spacing and the delay Doppler grid size of the OTFS.

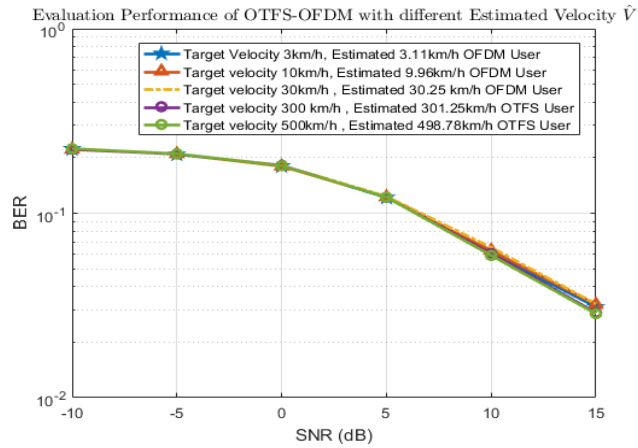

Figure 10. BERs Performance of OTFS-OFDM: The modulation scheme is 16-QAM

Moreover, the complexity of implementing a hybrid scheme is higher than that of using either OTFS or OFDM waveforms separated. This may be a concern in some practical applications. Overall, the results indicate that the hybrid scheme is a viable option for improving BER performance in high mobility scenarios, but careful design and implementation are crucial.

### 5.3. Complexity Analysis

In this part, we discuss the complexity analysis of the proposed Framework, a hybrid system that switches signal processing chains in the transmitter and receiver, facilitating the use of either OTFS or OFDM waveforms. A detailed analysis of the complexity of the MF-F algorithm is provided, divided into two distinct parts. In the first part, they focus on the MF step. This step includes a low-complexity circular shift operation with a complexity of $\mathcal{O}(M) + \mathcal{O}(N)$, giving a total complexity of $\mathcal{O}(MN)$. The second part, or step F, involves matrix calculations and has a complexity of $(\mathcal{O}(MN)^2)$ when the matrix operations are performed directly. Multiplication of the diagonal matrix has a complexity of $\mathcal{O}(MN)$ while the cyclic shift matrix operation has a complexity of $\mathcal{O}(MN)^2$. FFTs at point M and inverse FFTs at point N have respective complexities of $\mathcal{O}\big((MN)log_2(M)\big)$ and $\mathcal{O}\big((MN)log_2(N)\big)$. Consequently, the total complexity of the proposed MF-F algorithm is of the order of $\mathcal{O}((MN)log_2(MN))$. Table 2 shows various complexity parameters of the different algorithms used in different waveforms.

Table 2. Complexity Analysis.

| Waveform's | Algorithm | Complexity |
|---|---|---|
| OFDM [21] | FFT | $\mathcal{O}\big((N)log(N)\big)$ |
| OTFS [29], [33], [34] | SFFT | $\mathcal{O}\big((MN)log_2(MN)\big)$ |
| Proposed | FFT (if $\breve{v} \leq v_{threshold}$) | $\mathcal{O}\big((N)log(N)\big)$ |
| | SFFT (if $\breve{v} > v_{threshold}$) | $\mathcal{O}\big((MN)log_2(MN)\big)$ |

The aim of deploying the ISAC radar system for OTFS-OFDM is to improve the algorithm's performance in real-life situations. The integration of ISAC radar reduces the complexity of OTFS-OFDM implementation. Our proposed framework enables guided switching between OTFS and OFDM.This framework guarantees high data detection and low implementation complexity, which is particularly advantageous in high-mobility scenarios thanks to the MF-F algorithm. This approach improves system efficiency and reduces complexity. This approach improves system efficiency and reduces complexity. This integrated approach improves system efficiency, enables adaptation to changing channel conditions, improves the robustness of the communication system, and enhances data quality, robustness and mobility. Taking into account the complexity of OTFS and the challenge of high mobility OFDM, the proposed framework delivers a global and exhaustive solution. The integration of ISAC radar reduces the complexity of OTFS-OFDM implementation. The proposed framework enables guided switching between OTFS and OFDM, that facilitated by the ISAC radar. This framework guarantees high data detection and low implementation complexity, which is particularly advantageous in high mobility scenarios thanks to the MF-F algorithm. This approach improves system efficiency and reduces complexity. This integrated approach improves system efficiency, enables adaptation to changing channel conditions, improves the robustness of the communication system, and enhances data quality, robustness and mobility. By addressing the complexity of OTFS and the challenge of high mobility OFDM, the proposed framework provides a complete solution.

## 6. CONCLUSION

In this paper, we introduce a hybrid system that switches signal processing chains in the transmitter and receiver, facilitating the use of either OTFS or OFDM waveforms. This system is based on the integrating sensing and communication ISAC system, which employs velocity estimation. Our research has primarily concentrated on the study of OTFS, a waveform that is increasingly being adopted due to its responsiveness to high user mobility. We have put forth a selection strategy between OTFS and OFDM to better cater to user mobility. This strategy allows us to choose the most suitable approach based on the user's speed, utilizing the ISAC system, which offers superior estimation accuracy and low complexity. This study has oriented our observations to select one more convenient approach based on user's speed rate. The outcomes of our study have been highly gratifying, affirming the validity of our proposed concept. A significant benefit of our approach is its sustainability when a switching procedure is implemented in real world systems. More enhanced strategies could be suggested to preview, looking forward, we propose the development of more sophisticated strategies. These could encompass more adaptable, or even automated, switching procedures based on other criteria, which could be designed for immediate execution. To conclude, the switching selection strategy we propose serves as a potent instrument for enhancing the performance of OTFS-OFDM systems. Thereby presenting a promising direction for future research and development.

## CONFLICTS OF INTEREST

The authors declare no conflict of interest.

## AUTHORS

**Amina Darghouthi** was born in Tozeur Tunisia, in 1993. Doctoral student researcher in electrical engineering at the National Engineering School of Gabes (Tunisia).She is an esteemed member of the Research Laboratory Modeling, Analysis, and Control Systems (MACS), registered under LR16ES22 (www.macs.tn), actively involved in research. In addition to her research pursuits, Fatma is currently serving as a contractual lecturer at the National School of Engineers of Gabes, where she shares her knowledge and expertise with students.

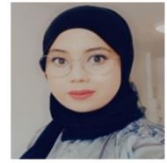

**Abdelhakim KHLIFI** is an assistant professor at the National Engineering School of Gabes, Tunisia. He received the Engineer degree from the Nation al Engineering School of Gabes in 2007, and the master's degree from the National Engineering School of Tunis in 2010, and the Ph.D. degree in 2015. 1. He specializes in signal processing and digital communications in his teaching endeavors. His main research activities focus on performances analysis of Waveform Optimization on 5G/6G systems.

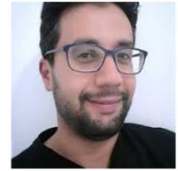

**HMAIED SHAIEK** is an associate professor at the National Conservatory of Arts and Crafts SITI School, France. He received the Engineer degree from the National Engineering School of Tunis in 2002, and the master's degree from the University de Bretagne Occidental in 2003, and the Ph.D. degree from the Lab-STICC CNRS Team, Telecom Bretagne, in 2007. He was with Canon Inc., until 2009 and left the industry to integrate with the National school of Ingenieurs de Brest, as a Lecturer, from 2009 to 2010. In 2011, He joined the CNAM, as an Associate Professor in electronics and signal proces sing. My teaching activities are in the fields of analog and digital electronics, microcontrollers programming, signal processing and digital communications. His main research activities focus on performances analysis of multicarrier modulations with nonlinear power amplifiers, PAPR reduction, and power amplifier linearization. He contributed to the FP7 EMPHATIC (www.ict-emphatic.eu/) European project and was involved in two national projects: ACCENT5 and WONG5 (www.wong5.fr), funded by the French National Research Agency.

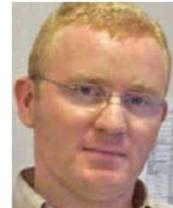

**Fatma Ben Salah** was born in Gafsa, Tunisia, in 1989. She earn ed her Bachelor's degree in Engineering in 2014 from the National School of Engineers of Gabes (Tunisia), specializing in Communication and Networking. Currently, Fatma is a doctoral student researcher in Electrical Engineering at the same institution. She is an esteemed member of the Research Laboratory Modeling, Analysis, and Control Systems (MACS), registered under LR16ES22 (www.macs.tn), actively involved in research. In addition to her research pursuits, Fatma is currently serving as a contractual lecturer at the National School of Engineers of Gabes, where she shares her knowledge and expertise with student

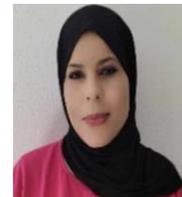

**RHAIMI Belgacem Chibani** is an Associate Professor in CSIE (Computer Sciences & Information Engineering). He joined the National Engineering High School at Gabes named (ENIG) where he is actually employed since Septemer1991. After a Doctorate Thesis earned at the National Engineering High School at Tunis (ENIT), he received the Ph.D. degree from ENIG, University of Gabes, Tunisia in 1992. He is a member of the Research Laboratory MACS at ENIG as activities supervisor dealing with Signal Processing and Communications Research field. Currently, his research areas cover Signal Processing and Mobile Communications. He is currently working with the University of Gabes. His research interests include Information and Signal Processing, Communications Engineering. He has published a number of papers on international regular organized conferences and journals (e.g., CESA, IFAC, autumn, spring, A2I and Summer Schools). He is a member of Communications Engineering staff at ENIG. He has been serving as a Program Committee Member dealing with Communications for a number of top national schools and activities.

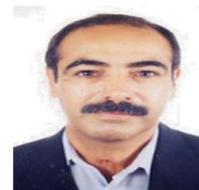